\documentclass[twocolumn,aps,prl,superscriptaddress]{revtex4-2}
\usepackage{subcaption}
\usepackage{graphicx}
\usepackage[version=3]{mhchem}
\usepackage{braket}
\usepackage{ulem}
\usepackage{here}
\usepackage{dcolumn}
\usepackage{color}
\usepackage{wrapfig}
\usepackage{bm}
\usepackage{xcolor}
\usepackage{lipsum}
\usepackage{cancel}
\usepackage{soul}
\usepackage{ragged2e}
\usepackage[utf8]{inputenc}
\usepackage{amsmath}
\usepackage[percent]{overpic}

\begin{document}
\title{XUV Transmission Spectroscopy Using a Tabletop High-Harmonic Source}
\author{Ryunosuke Takahashi}
\thanks{These authors contributed equally to this work.}
\email{bagdiners@gmail.com}
\affiliation{Department of Material Science, Graduate School of Science,
University of Hyogo, 3-2-1 Koto, Kamigori-cho, Ako-gun,
Hyogo 678-1297, Japan}

\author{Soudai Sakoda}
\thanks{These authors contributed equally to this work.}
\email{sakodasoudai@gmail.com}
\affiliation{Department of Material Science, Graduate School of Science,
University of Hyogo, 3-2-1 Koto, Kamigori-cho, Ako-gun,
Hyogo 678-1297, Japan}
\author{Kaede Yamada}
\affiliation{Department of Material Science, Graduate School of Science, University of Hyogo, 3-2-1 Koto, Kamigori-cho, Ako-gun, Hyogo 678-1297, Japan}
\author{Jumpei Horai}
\affiliation{Department of Material Science, Graduate School of Science, University of Hyogo, 3-2-1 Koto, Kamigori-cho, Ako-gun, Hyogo 678-1297, Japan}
\author{Shigetoshi Tomita}
\affiliation{Department of Material Science, Graduate School of Science, University of Hyogo, 3-2-1 Koto, Kamigori-cho, Ako-gun, Hyogo 678-1297, Japan}
\author{Yuto Shiokawa}
\affiliation{Department of Material Science, Graduate School of Science, University of Hyogo, 3-2-1 Koto, Kamigori-cho, Ako-gun, Hyogo 678-1297, Japan}
\author{Nobuhisa Ishii}
\affiliation{Kansai Institute for Photon Science, National Institutes for Quantum Science and Technology (QST), 8-1-7 Umemidai, Kizugawa, Kyoto 619-0215, Japan}
\author{Hiroki Wadati}
\affiliation{Department of Material Science, Graduate School of Science, University of Hyogo, 3-2-1 Koto, Kamigori-cho, Ako-gun, Hyogo 678-1297, Japan}
\affiliation{Institute of Laser Engineering, The University of Osaka, Suita, Osaka 565-0871, Japan}

\begin{abstract}
High-harmonic generation (HHG) provides a coherent ultrashort-pulse light source in the extreme-ultraviolet (XUV) region and has potential applications in time-resolved spectroscopy and materials characterization. In this study, we constructed an HHG system using a Yb:KGW laser system (PHAROS, Light Conversion) as the driving source and Ar gas as the nonlinear medium. High-order harmonics were observed up to a photon energy of 70.6~eV, corresponding to the 59th harmonic order. As an application of the developed XUV source, we measured the transmission of a Mg thin film and a $\mathrm{Si_3N_4}$ membrane. In addition, the surface chemical state of the Mg thin film was characterized by X-ray photoelectron spectroscopy (XPS), and the results were compared with the transmission measured using the HHG source. This comparison was used to examine the relationship between the surface condition, including surface oxidation, and the XUV transmission of the Mg thin film. These results demonstrate the applicability of the developed Ar-based HHG source to the characterization of thin-film transmission in the XUV spectral region.
\end{abstract}

\maketitle
High-order harmonic generation (HHG) in gases provides a compact and coherent source of extreme-ultraviolet (XUV) and soft-X-ray radiation
\cite{Corkum1993HHG,Lewenstein1994HHG,Krausz2009Attosecond,
Chen2010HHG,Popmintchev2012HHG,Lorek2014HighRepetitionHHG,Ishii2012WaterWindowHHG,
Ishii2014CEPHHG}
and has enabled a broad range of applications, including ultrafast
spectroscopy
\cite{Goulielmakis2010Electron,Pertot2017TRXAS},
element-selective measurements
\cite{LaOVorakiat2009Demagnetization,Mathias2012Exchange,
Willems2015HHGMCD,Willems2020OISTR,
vonKorffSchmising2020ElementSpecific,Tsai2026}, and coherent imaging
\cite{Sandberg2007Imaging,Gardner2017Imaging}. 
Beyond time-resolved measurements, broadband HHG sources are also
well suited to static spectroscopic applications, including XUV and soft-x-ray transmission spectroscopy
\cite{Buades2018XAFS,Popmintchev2018XAFS} and X-ray absorption near-edge spectroscopy \cite{Ishii2026OxygenKXANES}.
Reference spectra and noise-cancellation schemes improve the
stability and reliability of XUV spectroscopic measurements
\cite{Yao2020TabletopXUV,Johnsen2023NoiseCanceledXUV}. By normalizing the spectrum transmitted through a sample to a reference spectrum recorded without the sample, the energy-dependent transmittance spectrum can be obtained over the bandwidth of the HHG source \cite{Schmidt2018OxygenK}.

\begin{figure*}[t]
    \centering
    \includegraphics[width=\linewidth]{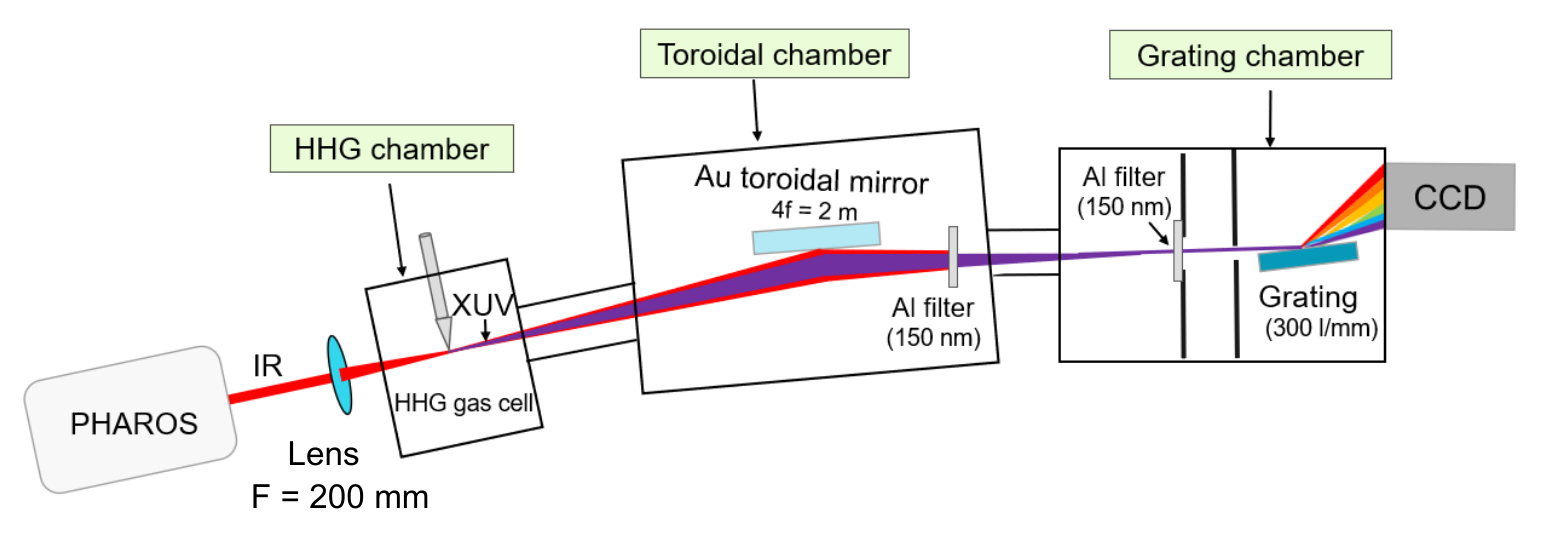}
    \caption{\justifying Schematic illustration of the high-harmonic generation (HHG) spectroscopy setup. An intense infrared laser pulse is focused into an Ar gas jet to generate high harmonics. The residual fundamental beam and low-order harmonics were removed using two 150 nm Al filters. The HHG beam was then focused onto the sample by an Au toroidal mirror. The transmitted harmonics were dispersed by a diffraction grating and detected with a CCD camera.}
    \label{fig:device}
\end{figure*}

XUV transmission measurements are highly sensitive to the composition and thickness of thin films because the attenuation lengths of many materials typically range from a few nanometers to several hundred nanometers, depending on the material and photon energy \cite{Henke1993XRay}. Tabulated atomic scattering factors and optical constants, such as those provided by the Center for X-Ray Optics (CXRO), are therefore widely used to calculate the transmission of thin films and filters \cite{Henke1993XRay}. Agreement between measured and calculated transmission spectra, particularly in the positions of well-defined absorption edges and in the overall spectral shape, provides a practical check of the photon-energy calibration and performance of an XUV spectrometer \cite{Burcklen2024Al}. However, actual thin-film samples do not always behave as ideal homogeneous layers with their nominal composition, mass density, and thickness \cite{Windt1988ThinFilms,Burcklen2024Al}. Surface oxidation, contamination, thickness nonuniformity, and deviations from the bulk density can modify their XUV optical response. Consequently, measured transmission spectra may exhibit systematic deviations from calculations based on bulk optical constants and nominal film parameters \cite{Windt1988ThinFilms,Burcklen2024Al}.

This issue is particularly relevant for metallic Mg films. Magnesium is highly reactive under ambient conditions and forms a surface layer containing MgO and, depending on the exposure conditions, hydroxide and carbonate species
\cite{Fotea2006MgSurface,Burke2012MgXPS}.
The thickness and chemical composition of this surface layer can significantly affect XUV transmission because the optical responses of metallic Mg and MgO differ substantially in the XUV region
\cite{VidalDasilva2010Mg,Hanson1972MgO}.
Consequently, a comparison with the calculated transmission of pure Mg may not adequately describe the measured spectrum of an air-exposed Mg film.
X-ray photoelectron spectroscopy (XPS), which is sensitive to both elemental composition and chemical state within the near-surface region, provides complementary information for evaluating such surface oxidation
\cite{Fotea2006MgSurface,Burke2012MgXPS}.

In this study, we constructed a gas-phase HHG setup and employed the generated XUV radiation for transmission measurements of commercially available thin films with known nominal thicknesses. The transmission spectra of Si$_3$N$_4$ membranes and Mg films were measured and compared with calculations based on the CXRO optical-constant database. A commercially available $\mathrm{Si_3N_4}$ membrane TEM window grid with a nominal membrane thickness of 50~nm, a $0.5 \times 0.5~\mathrm{mm^2}$ window, and a 200-$\mu$m-thick frame (Alliance Biosystems) was used to examine whether the measured attenuation followed the expected thickness dependence.

For the Mg film, the measured transmission was further analyzed by considering a layered model containing metallic Mg and an oxidized surface layer. The surface electronic and chemical states were independently investigated by XPS.
XPS measurements were performed using a PHI 5000 VersaProbe III spectrometer (ULVAC-PHI) equipped with a monochromated Al K$\alpha$ source ($h\nu = 1486.6$ eV). Survey and high-resolution spectra were acquired using analyzer pass energies of 280 and 55 eV, respectively. Photoelectrons were collected in the normal-emission geometry, corresponding to an emission angle of $0^\circ$ relative to the surface normal ($90^\circ$ relative to the sample surface). The pressure in the analysis chamber was in the $10^{-8}$ Pa range, and the overall energy resolution was approximately 0.5 eV.
The binding-energy scales of the C 1$s$ and O 1$s$ spectra were corrected by setting the adventitious C–C/C–H component of the C 1$s$ spectrum to 284.8 eV. This procedure corresponded to a shift of $-1.79$ eV from the as-measured energy scale, and the same shift was applied to the O 1$s$ spectrum. In contrast, the Mg 2$p$ spectrum is presented on the as-measured energy scale because the metallic Mg$^0$ component already appears near the expected binding energy. Applying the C 1$s$-derived shift to the Mg 2$p$ spectrum would move the metallic component from 49.68 to 47.89 eV, which is physically unreasonable. This behavior is consistent with differential charging between the conducting Mg layer and the insulating surface products.

By combining HHG-based XUV transmission measurements, optical calculations, and XPS, we examine how surface oxidation contributes to the transmission spectrum of a nominally metallic Mg film. The present work demonstrates a laboratory-scale procedure for validating XUV thin-film transmission measurements and for identifying deviations from ideal-film calculations arising from the actual chemical condition of the sample.

A schematic diagram of the experimental setup is shown in Figure~\ref{fig:device}. The tabletop HHG source was driven by infrared laser pulses with a duration of $\sim$ 200 fs, a repetition rate of $1$~kHz, and an average power of $1$~W. The infrared beam was focused into an Ar gas jet using a lens with a focal length of $200$~mm. Gas nozzles with inner diameters of $0.2$ and $0.3$~mm were used, and the Ar backing pressure was maintained at approximately $0.05$~MPa. The operating pressure in the HHG generation chamber was in the range of $2$--$5$~Pa.

The residual fundamental radiation and low-order harmonics were suppressed using two $150$-nm-thick Al filters. Because the filter transmittance decreases sharply near the Al $L$ absorption edge, the 61st and higher harmonics were not observed in the detected HHG spectrum. The transmitted XUV beam was focused onto the sample using an Au-coated toroidal mirror. 
A knife-edge measurement of the 31st
harmonic yielded a transverse full width at half maximum of approximately 340~$\mu$m at the sample position
(see Fig.~S1 of the Supplemental Material).
After transmission through the sample, the high-order harmonics were spectrally dispersed using a diffraction grating with 300 grooves mm$^{-1}$ (30-006, Shimadzu) and detected using a Peltier-cooled X-ray CCD camera (Newton DO920P-BEN, Andor).
For each measurement, a background spectrum was recorded under otherwise identical conditions with the Ar gas supply turned off and was subtracted from the measured spectrum.

\begin{figure}[t]
\centering
\includegraphics[width=\linewidth]{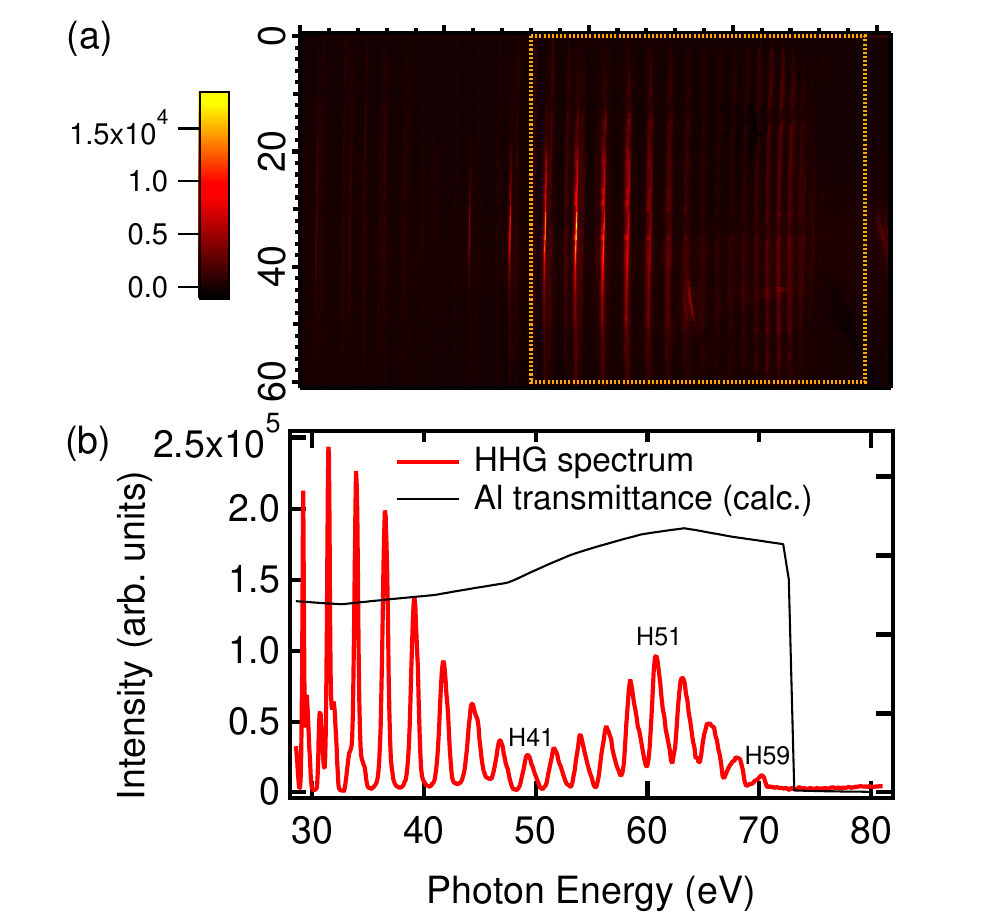}
\caption{\justifying
High-order harmonic emission generated by the tabletop HHG source.
(a) Raw CCD image of the spectrally dispersed HHG emission.
The orange dotted rectangle indicates the region used for spectral extraction.
(b) HHG spectrum obtained by integrating the CCD counts within the dotted rectangular region in (a) along the vertical direction at each horizontal position and calibrating the horizontal detector coordinate to photon energy.
The black curve shows the calculated Al transmittance normalized to its maximum value and plotted on an arbitrary vertical scale to facilitate comparison of its spectral shape near the Al $L_{2,3}$ absorption edge.
}
\label{fig:HHG_spectra}
\end{figure}

\begin{figure}[t]
\centering
\includegraphics[width=\linewidth]{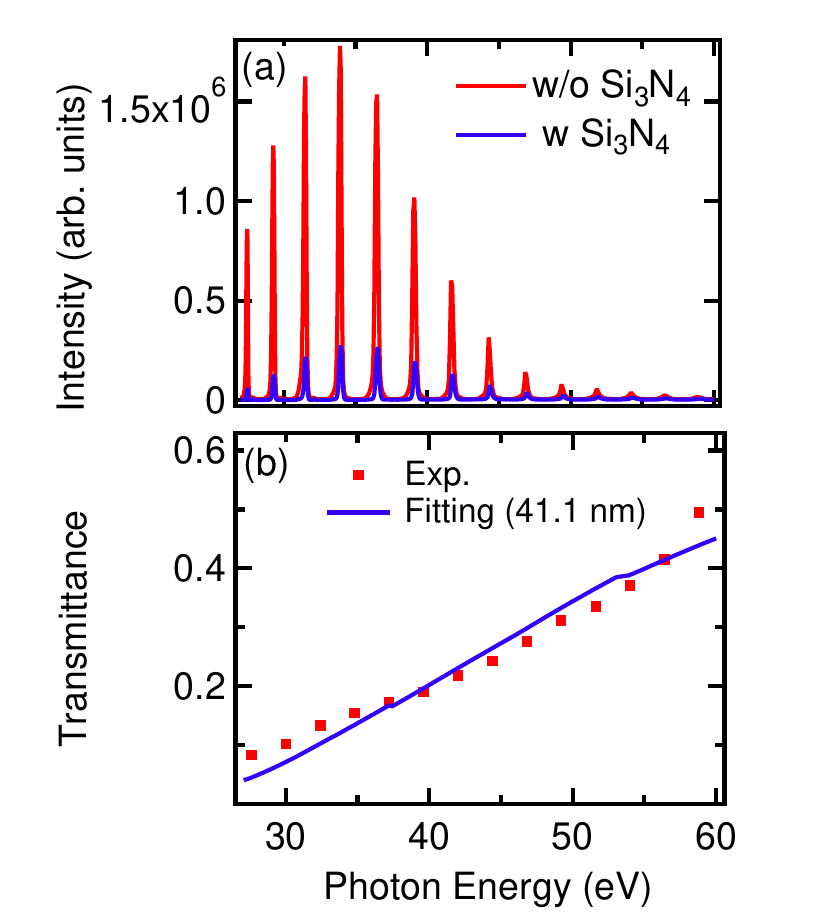}
\caption{\justifying
(a) High-harmonic spectra measured without (red) and with (blue) a
commercially available Si$_3$N$_4$ membrane (nominal thickness: 50~nm).
(b) Experimental transmittance of the Si$_3$N$_4$ membrane and the CXRO calculation for ($d_{\mathrm{eff}}=41.1\pm0.9$) nm.
}
\label{fig:SiN_transmission}
\end{figure}
The central wavelength of the fundamental infrared laser pulses was measured using a spectrometer and determined to be $1035.9$~nm, corresponding to a fundamental photon energy of approximately $1.197$~eV. The measured spectrum of the fundamental laser pulses is shown in
Fig.~S2 of the Supplemental Material. The photon energies of the observed odd-order harmonics were assigned using $E_n=nE_{\mathrm{fund}}$. Because the relationship between the horizontal CCD coordinate and photon energy varied slightly between measurement series, the photon-energy calibration was performed independently for each series. The CCD positions of the harmonic peaks in the corresponding reference spectrum were fitted to their assigned photon energies using a second-order polynomial. When clearly resolved, the Mg and Al $L_{2,3}$ absorption edges were included as additional calibration points. The same calibration was applied to the sample and reference spectra acquired without changing the spectrometer configuration. Details of the calibration procedure are provided in the Supplementary Material.
\begin{figure}[t]
\centering
\includegraphics[width=0.9\linewidth]{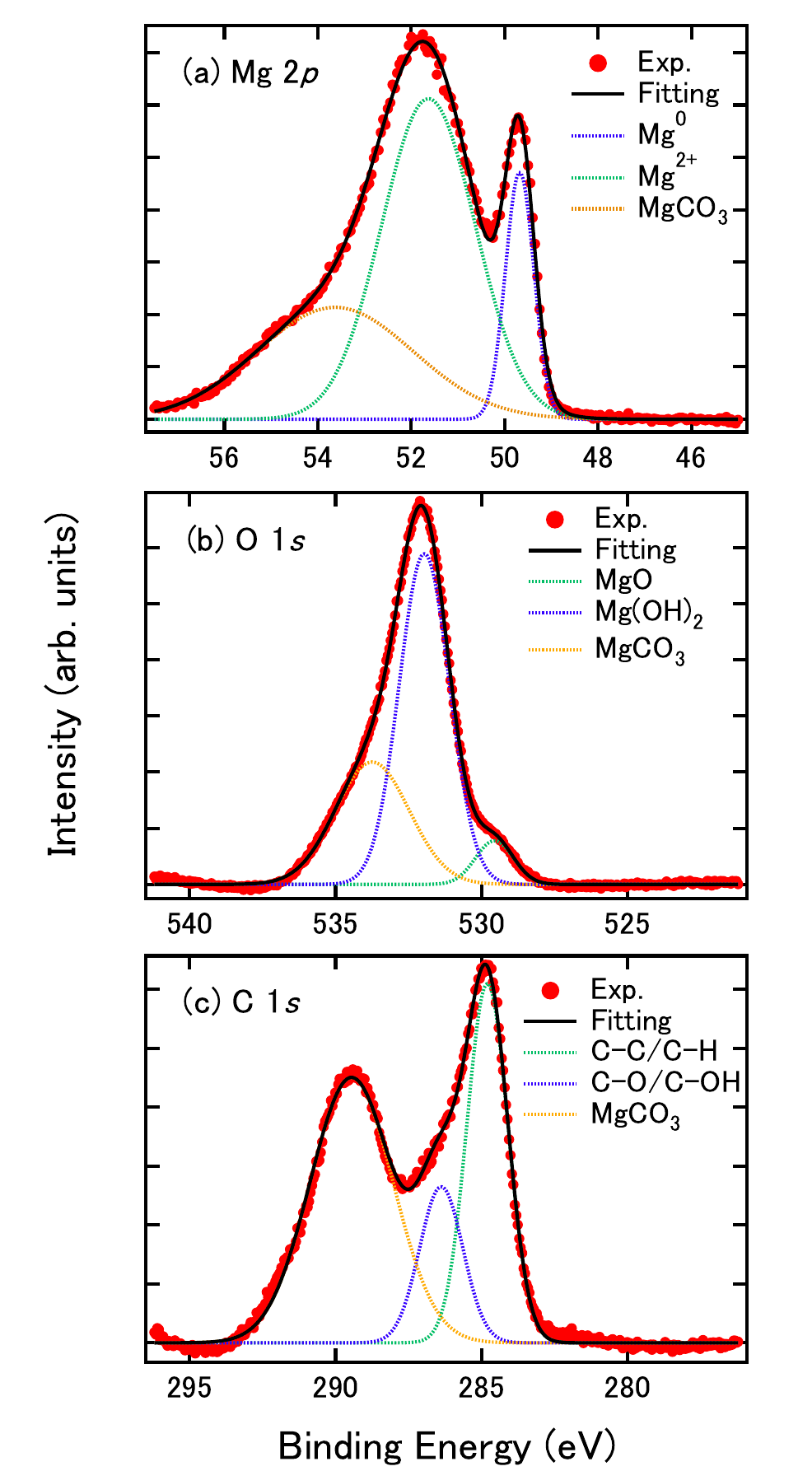}
\caption{\justifying
XPS core-level spectra of the Mg thin film in the (a) Mg~$2p$, (b) O~$1s$, and (c) C~$1s$ regions. Red circles show the experimental spectra, black solid lines show the total fits, and dotted lines show the individual components. The Mg~$2p$ spectrum was decomposed into metallic Mg$^{0}$, Mg$^{2+}$, and carbonate-related Mg components; the O~$1s$ spectrum into MgO, Mg(OH)$_2$, and MgCO$_3$ components; and the C~$1s$ spectrum into C--C/C--H, C--O/C--OH, and carbonate components. Detailed fitting conditions and parameters are provided in the Supplementary Material. The C 1s and O 1s binding-energy axes are shown after applying the
$-1.79$~eV correction, whereas the Mg 2p axis is shown on the as-measured scale.
}
\label{fig}
\end{figure}

Figure~\ref{fig:HHG_spectra} shows the high-order harmonic spectrum generated by the tabletop HHG source. The source was driven by laser pulses with a duration of $182$~fs, a repetition rate of $1$~kHz, and an average power of $1$~W. The laser pulses were focused into an Ar gas jet emitted from a nozzle with a diameter of $0.2$~mm. The Ar backing pressure at the gas inlet valve was $0.05$~MPa, while the pressure in the HHG chamber was maintained at approximately $2$~Pa during data acquisition. Each CCD image was acquired with an exposure time of $2$~s, and 50 images were accumulated, corresponding to a total acquisition time of $100$~s. High-order harmonics were observed up to the 59th order, corresponding to $70.6$~eV and labeled H59 in Figure~\ref{fig:HHG_spectra}. Below approximately $35$~eV, additional peaks were visible adjacent to the main HHG peaks. These features, appearing at positions corresponding to the apparent 29.5th, 28.5th, and lower orders, originated from second-order diffraction by the grating. Among the observed harmonics, the 29th-order harmonic exhibited the highest intensity.

Using this HHG source, transmission measurements were carried out for
commercially available Si$_3$N$_4$ membranes, and Mg thin films with known nominal
thicknesses. The transmittance of each sample was determined from the ratio of
the harmonic intensities measured with and without the sample. The measured
transmittance was compared with calculations based on the optical constants
provided by CXRO to evaluate the reliability of
the present measurement system. The following paragraphs describe the results for each sample.

First, we measured the transmittance of a commercially available Si$_3$N$_4$ membrane with a manufacturer-specified nominal thickness of 50~nm. Figure~\ref{fig:SiN_transmission}(a) compares the HHG
spectra measured without and with the Si$_3$N$_4$ membrane. 
Each CCD image was acquired with an exposure time of $0.3$~s, and 50 images were accumulated, corresponding to a total acquisition time of $15$~s.
A gas nozzle with inner diameters of $0.3$~mm was used, and the Ar backing pressure was maintained at approximately $0.05$~MPa. The operating pressure in the HHG generation chamber was $\sim$$3$~Pa.
The pulse duration was $\sim$ 200 fs.
Upon insertion of
the membrane, the intensity of each harmonic peak decreased, indicating the attenuation of the XUV radiation by
the membrane.

The transmittance was obtained from the intensity ratio of the corresponding
harmonic peaks measured with and without the membrane. The resulting
transmittance is shown in Fig.~\ref{fig:SiN_transmission}(b), together with
the transmittance calculated using the CXRO optical constants. Fitting the transmittance spectrum yielded an effective film thickness of \(d_{\mathrm{eff}} = 41.1 \pm 0.9\) nm, where the uncertainty represents one standard deviation estimated from the fitting residuals. The corresponding 95\% confidence interval was 39.2–43.2 nm. Details of the fitting procedure are provided in Table S1 of the Supplemental Material.
Although this value is 
smaller than the manufacturer's nominal thickness of 50~nm, the calculated
spectrum successfully reproduces the overall photon-energy dependence of the
measured transmittance. The difference between the nominal and calculated
thicknesses may arise from uncertainties in the nominal membrane thickness,
material density, membrane composition, or experimental uncertainties in the
transmission measurement. Nevertheless, the overall agreement demonstrates
that the present HHG-based measurement system reproduces the overall spectral dependence of the XUV transmittance of thin-film samples.

Since Mg is readily oxidized, an oxide layer that forms at the surface may
influence its optical response. Therefore, prior to comparing the
transmittance measured using HHG with the
calculated values, the surface chemical state of the sample was examined
by XPS.

As shown in Fig.~\ref{fig}(a), the Mg~$2p$ spectrum retains a distinct Mg$^{0}$ contribution, indicating that metallic Mg remains beneath the reacted surface region. The dominant Mg$^{2+}$ contribution is consistent with the oxidation and hydroxylation of the surface. Although the higher-binding-energy component may contain contributions from carbonate-related Mg species, its assignment based solely on the Mg~$2p$ spectrum is not unambiguous due to the substantial overlap among the Mg~$2p$ signals of MgO, Mg(OH)$_2$, and MgCO$_3$\cite{Skaanvik2025}. The O~$1s$ spectrum [Fig.~\ref{fig}(b)] can be decomposed into a major hydroxide- or hydrated-Mg-related component and smaller MgO- and carbonate-related contributions. Consistently, the C~$1s$ spectrum [Fig.~\ref{fig}(c)] contains a clear carbonate-related component in addition to adventitious hydrocarbon and oxygenated-carbon species. Taken together, these results indicate that the air-exposed Mg film is covered by a surface reaction region containing MgO, Mg(OH)$_2$, and carbonate-related species, along with adsorbed carbonaceous species. Because the Mg$^{2+}$ components strongly overlap and the depth distribution of the constituent phases is not independently known, the thicknesses of the individual oxide, hydroxide, and carbonate regions cannot be determined uniquely from the present XPS spectra. Nevertheless, the XPS results demonstrate that the optical response cannot be described solely in terms of an ideal metallic Mg film, and that the reacted surface region should be considered when comparing the measured HHG transmittance with the calculated values. Additional XPS spectra are shown in Fig.~S3 of the Supplemental
Material.

\begin{figure}[t]
\centering
\includegraphics[width=\linewidth]{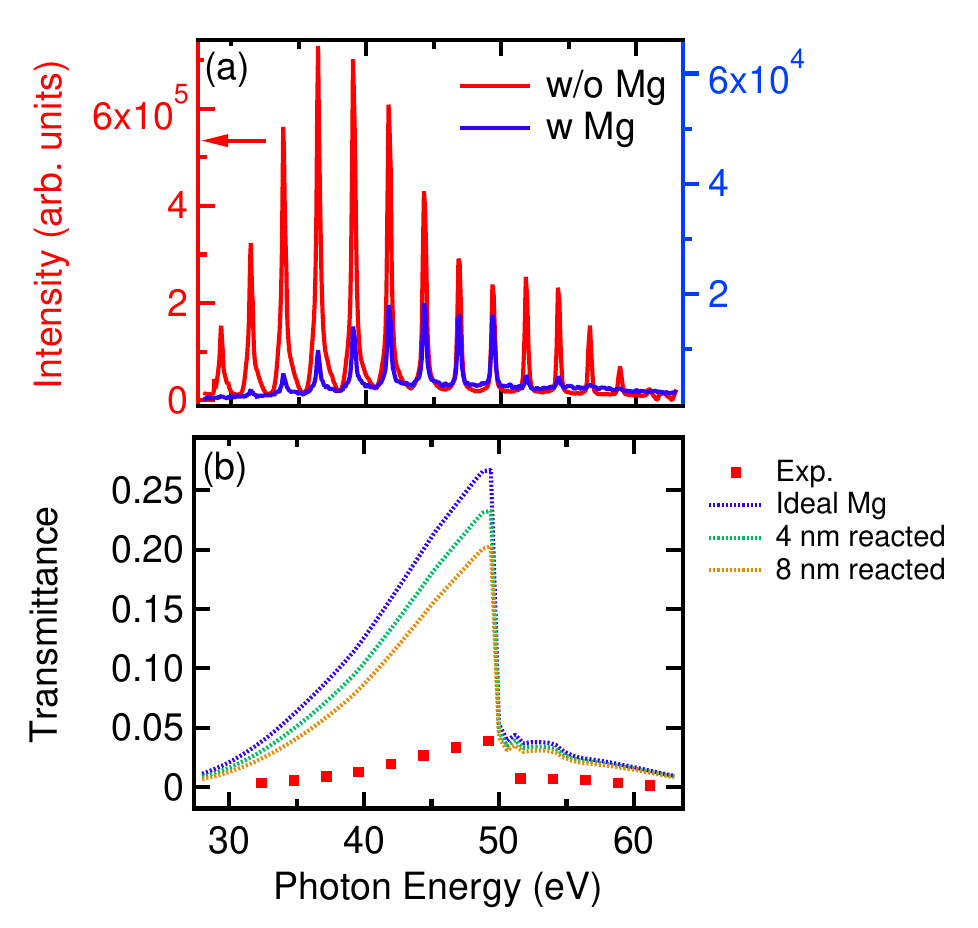}
\caption{\justifying
(a) High-order harmonic spectra measured with (blue) and without (red) the Mg/Parylene-N sample. For each spectrum, the background acquired without Ar gas was subtracted. The transmittance was obtained from the ratio of the two background-subtracted spectra.
(b) Experimental transmittance of the Mg/Parylene-N sample and CXRO-based calculations for an ideal 200-nm-thick Mg layer and XPS-informed surface-reaction layers with thicknesses of 4 and 8~nm. The total Mg-derived thickness was fixed at 200~nm, and a 100-nm-thick Parylene-N supporting layer was included in all calculations.
}
\label{fig:MgHHG}
\end{figure}
Figure~\ref{fig:MgHHG}(a) shows the high-order harmonic spectra acquired
with and without the Mg thin film.
Each CCD image was acquired with an exposure time of $1$~s, and 20 images were accumulated, corresponding to a total acquisition time of $20$~s.
A gas nozzle with inner diameters of $0.3$~mm was used, and the Ar backing pressure was maintained at approximately $0.05$~MPa. The operating pressure in the HHG generation chamber was $\sim$$5$~Pa.
The pulse duration was $\sim$ 200 fs.
The insertion of the Mg thin film
substantially reduced the harmonic intensity over the measured
photon-energy range. The experimental transmittance was determined by
taking the intensity ratio of the spectra measured with and without the
film. Figure~\ref{fig:MgHHG}(b) compares the resulting transmittance
spectrum with the calculated spectrum based on the optical constants
obtained from the CXRO database. 

To obtain an order-of-magnitude estimate of the thickness of the reacted surface region, the attenuation of the metallic Mg~$2p$ component was considered using a planar overlayer model. The observed reacted-Mg/metallic-Mg Mg 2p area ratio, together with a practical effective attenuation length of 2--3~nm, gives an effective reacted-layer thickness of approximately 4--8~nm, depending on the assumed composition and photoelectron emission angle. Because the Mg~$2p$ signals of the individual Mg$^{2+}$ compounds strongly overlap and their depth distribution is not known independently, this range should not be regarded as a quantitative determination of the thicknesses of the individual chemical phases. Instead, 4 and 8~nm were used as lower and upper bounds in the optical calculations.

For the transmittance calculations, an XPS-informed nominal composition of the reacted region was estimated from the O~$1s$ component areas after accounting for the number of oxygen atoms per Mg atom and the molar volumes of MgO, Mg(OH)$_2$, and MgCO$_3$. This procedure gives nominal relative volume fractions of approximately 0.06, 0.66, and 0.28, respectively. The total geometrical thickness of the Mg-derived layer was fixed at 200~nm, such that the 4-nm model consisted of 196~nm of metallic Mg, 0.24~nm of MgO, 2.65~nm of Mg(OH)$_2$, and 1.11~nm of MgCO$_3$. The corresponding thicknesses for the 8-nm model were 192, 0.48, 5.30, and 2.22~nm, respectively. The 100-nm-thick Parylene-N supporting membrane was included in all calculations. The transmittance of each constituent was calculated using the optical constants tabulated in the CXRO database \cite{Henke1993XRay}, and the total transmittance was obtained by multiplying the individual contributions within the Beer--Lambert approximation. No parameters were adjusted to reproduce the measured HHG transmittance.

At the experimental peak energy of 49.2~eV, the measured transmittance was 0.0389, whereas the calculated values were approximately 0.267, 0.232, and 0.202 for the ideal Mg film and the 4- and 8-nm reacted-layer models, respectively. The reacted-layer models therefore reduce the calculated transmittance by approximately 13--24\% relative to the ideal-film calculation, but the calculated values remain substantially larger than the experimental value. These results indicate that the surface reaction layer alone cannot account for the low measured transmittance. Additional factors, including the effective optical thickness of the sample, the finite spectral bandwidth and energy calibration of the individual harmonics, and the experimental normalization, may also contribute to the discrepancy. Thus, the XPS-informed calculations should be regarded as a sensitivity analysis of the influence of the reacted surface region rather than a quantitative reproduction of the measured HHG transmittance.

\section{Conclusion}

In conclusion, we constructed a tabletop high-harmonic source driven by a Yb:KGW laser and generated high-order harmonics up to the 59th order, corresponding to a photon energy of 70.6~eV. The applicability of the source to XUV transmission measurements was examined using a Si$_3$N$_4$ membrane and an Mg/Parylene-N sample. The measured transmittance of the Si$_3$N$_4$ membrane was well reproduced by a model based on the CXRO optical constants. The fit yielded an effective thickness of \(d_{\mathrm{eff}}=41.1\pm0.9\) nm, where the uncertainty represents one standard deviation estimated from the fitting residuals.

For the Mg sample, XPS measurements revealed that metallic Mg remained beneath a reacted surface region containing MgO, Mg(OH)$_2$, and carbonate-related species. A semiquantitative analysis of the attenuation of the metallic Mg~$2p$ component suggested an effective total reacted-layer thickness of approximately 4--8~nm, although the thicknesses and depth distributions of the individual chemical phases could not be determined uniquely. XPS-informed optical models using 4- and 8-nm reacted layers reduced the calculated XUV transmittance relative to that of an ideal metallic Mg film. However, the calculated transmittance remained substantially higher than the experimental values and did not fully reproduce their photon-energy dependence. The reacted surface region therefore cannot alone account for the discrepancy. Variations in the actual Mg and Parylene-N thicknesses and densities, lateral nonuniformity, uncertainties in the optical constants, the finite spectral bandwidth and energy calibration of the harmonics, and the experimental normalization may also contribute.

These results demonstrate the feasibility of the developed HHG source for laboratory-scale XUV transmission measurements. They also highlight the importance of independent chemical and structural characterization, together with careful consideration of the experimental spectral response, when interpreting the XUV transmittance of air-sensitive thin films.

\section{acknowledgments}
This work was supported by JSPS KAKENHI under Grant
Nos.~JP19H02623, JP23H01108, JP23K25805, JP24K01390,
JP25H00864, and JP25H01251, by the MEXT Quantum Leap
Flagship Program (MEXT Q-LEAP) under Grant
No.~JPMXS0118068681, and by JST PRESTO under Grant
No.~JPMJPR2002. R.T. acknowledges support from JSPS
KAKENHI under Grant No.~JP23KJ1854. Additional support
was provided by the Asahi Glass Foundation.

\section*{Data Availability}
The data that support the findings of this article are available from
the authors upon reasonable request.

\bibliography{references}

\end{document}


\title{Supplementary Material for ``XUV Transmission Spectroscopy Using a Tabletop High-Harmonic Source''}

\author{Ryunosuke Takahashi, Soudai Sakoda, Kaede Yamada, Jumpei Horai, Shigetoshi Tomita, \\ Yuto Shiokawa, 
Nobuhisa Ishii, and Hiroki Wadati}

\maketitle

\section{Transverse beam-size characterization}

The transverse beam size of the high-order harmonics was evaluated using the knife-edge method. For this measurement, the pulse duration, repetition rate, and average power of the driving laser were $182$~fs, $1$~kHz, and $1$~W, respectively. The Ar backing pressure was $0.05$~MPa, and a gas nozzle with an inner diameter of $0.2$~mm was used. The pressure in the HHG generation chamber during operation was approximately $2$~Pa.

\begin{figure}[H]
  \centering
  \includegraphics[width=0.8\linewidth]{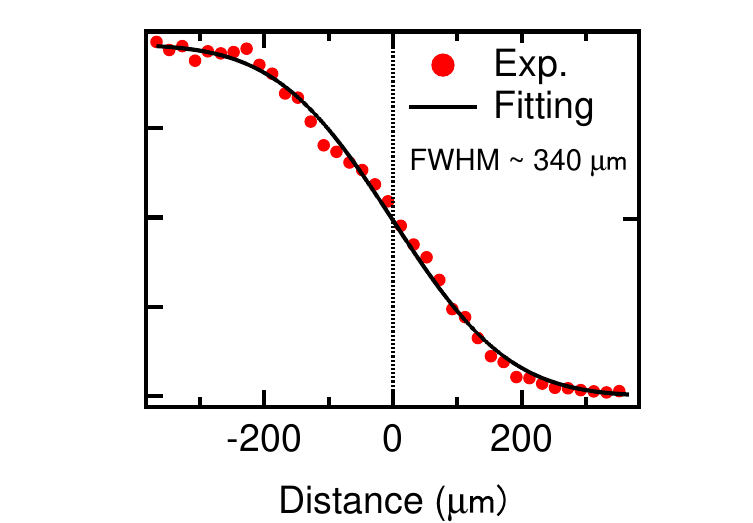}
  \caption{Knife-edge scan of the 31st harmonic. The solid curve represents the error-function fit to the experimental data, and the dotted curve shows the corresponding transverse intensity profile obtained by differentiating the fitted curve. The estimated transverse FWHM along the scan direction was approximately $340$~$\mu$m in full width at half maximum.}
  \label{fig:supp-beam-size}
\end{figure}

Figure~\ref{fig:supp-beam-size} shows the intensity of the 31st harmonic as a function of the knife-edge position. The measured knife-edge curve was fitted using
\begin{equation}
I(x)=I_{\mathrm{bg}}+\frac{I_0}{2}
\left[1+\operatorname{erf}\left(\frac{x-x_0}{\sqrt{2}\sigma}\right)\right],
\label{eq:knife-edge}
\end{equation}
where $I_{\mathrm{bg}}$ is the background intensity, $I_0$ is the integrated beam intensity, $x_0$ is the beam-center position, and $\sigma$ is the standard deviation of the corresponding Gaussian intensity profile. Differentiation of Eq.~\eqref{eq:knife-edge} gives the transverse beam profile. Its full width at half maximum,
$2\sqrt{2\ln 2}\,\sigma$, was approximately $340$~$\mu$m.

\section{Photon-energy calibration}

\begin{figure}[H]
  \centering
  \includegraphics[width=0.8\linewidth]{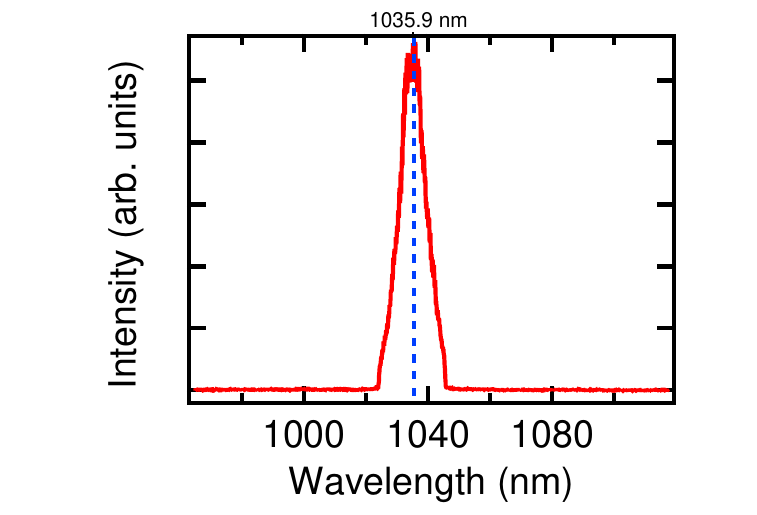}
  \caption{Spectrum of the fundamental infrared laser pulses. The peak wavelength was $1035.9$~nm.}
  \label{fig:supp-fundamental}
\end{figure}

The spectrum of the fundamental infrared laser pulses was measured
using a spectrometer. As shown in
Figure~\ref{fig:supp-fundamental}, the peak wavelength was
$1035.9$~nm. The corresponding fundamental photon energy was
calculated as

\begin{equation}
E_{\mathrm{fund}}
=
\frac{hc}{\lambda_{\mathrm{fund}}}
=
\frac{1239.84}{1035.9}
=
1.1969~\mathrm{eV}.
\label{eq:fundamental-energy}
\end{equation}

The photon energy of a harmonic of order $n$ was therefore
assigned as

\begin{equation}
E_n=nE_{\mathrm{fund}}.
\label{eq:harmonic-energy}
\end{equation}

Because the relationship between the horizontal CCD coordinate
and photon energy varied slightly between measurement series,
the photon-energy calibration was performed independently for
each measurement series rather than using a single fixed
pixel-to-energy conversion.

For each measurement series, the horizontal CCD positions of the
observed odd-order harmonic peaks were determined from the
corresponding reference HHG spectrum, and their photon energies
were assigned using Eq.~\eqref{eq:harmonic-energy}. When clearly
resolved in the corresponding spectra, the Mg and Al
$L_{2,3}$ absorption edges at $49.5$ and $72.7$~eV, respectively,
were included as additional calibration points \cite{Bearden1967}.

Because the relationship between the horizontal CCD coordinate
and photon energy was not strictly linear, it was represented by
a second-order polynomial,

\begin{equation}
E(P)=aP^2+bP+c,
\label{eq:energy-calibration}
\end{equation}

where $P$ is the horizontal CCD coordinate. The coefficients
$a$, $b$, and $c$ were determined separately for each measurement
series by an unweighted least-squares fit to the available
harmonic-peak and absorption-edge calibration points. The same
pixel-to-energy conversion was applied to the sample and reference
spectra acquired within the same measurement series without
changing the spectrometer configuration. Therefore, the absolute
CCD positions and polynomial coefficients were not common to all
measurements and are not reported as a single universal
calibration function.

\section{Determination of thin-film transmittance}

For each harmonic order, the sample transmittance was determined from the background-subtracted integrated intensities according to
\begin{equation}
T(E_n)=\frac{I_{\mathrm{sample}}(E_n)-I_{\mathrm{sample,bg}}(E_n)}
{I_{\mathrm{ref}}(E_n)-I_{\mathrm{ref,bg}}(E_n)},
\label{eq:experimental-transmittance}
\end{equation}
where $I_{\mathrm{sample}}$ and $I_{\mathrm{ref}}$ denote the integrated harmonic intensities acquired with and without the sample, respectively. The same integration window was applied to the sample and reference spectra for each harmonic order. The background spectra were acquired under otherwise identical conditions with the Ar gas supply turned off.

\section{Effective thickness of the Si$_3$N$_4$ membrane}

The nominal thickness of the commercially available Si$_3$N$_4$
membrane was $50$~nm. The calculated transmittance was obtained using
the CXRO optical constants \cite{Henke1993}, with the density fixed at
$3.44$~g~cm$^{-3}$, the tabulated value for Si$_3$N$_4$ in the CXRO
database \cite{CXRODensity}. Because the absorption-based transmittance
primarily constrains the areal density $\rho d$, the density and
thickness were not treated as independent fitting parameters.

Starting from the CXRO transmittance $T_{50}(E)$ calculated for a
$50$-nm-thick membrane, the transmittance for an effective thickness
$d$ was evaluated as
\begin{equation}
T(E;d)=\left[T_{50}(E)\right]^{d/(50~\mathrm{nm})}.
\label{eq:thickness-scaling}
\end{equation}
An unweighted least-squares fit in transmittance space over
$25.2$--$58.8$~eV, containing 15 experimental points, yielded
$d_{\mathrm{eff}}=41.1\pm0.9$~nm. The uncertainty represents one
standard deviation estimated from the fitting residuals, and the
corresponding 95
statistical uncertainty does not include possible systematic
uncertainties in the CXRO optical constants or the assumed density.

Restricting the fit to the higher-signal region from
$25.2$--$44.4$~eV, containing nine points, gave
$d_{\mathrm{eff}}=40.3$~nm, which is consistent with the full-range
result within its uncertainty. This agreement supports the robustness
of the fitted effective thickness against the choice of fitting range.

\begin{table}[H]
\centering
\caption{Results of the Si$_3$N$_4$ effective-thickness analysis.
The density was fixed at $3.44$~g~cm$^{-3}$ for both fits. The
uncertainty reported for the primary fit is the residual-based
$1\sigma$ statistical uncertainty.}
\label{tab:supp-si3n4-fit}
\begin{tabular}{lcccc}
  \hline
  Fit & Range (eV) & Points & $d_{\mathrm{eff}}$ (nm) & RMSE \\
  \hline
  Primary & 25.2--58.8 & 15 & $41.1\pm0.9$ & 0.0257 \\
  Restricted (high signal) & 25.2--44.4 & 9 & 40.3 & 0.0220 \\
  \hline
\end{tabular}
\end{table}

\section{Details of the XPS peak fitting}
\label{sec:xps_fitting}

The Mg~$2p$, O~$1s$, and C~$1s$ spectra were fitted over the ranges 45.0--57.5, 523--543, and 278--298~eV, respectively. Each spectrum was described by three Gaussian components and a linear background. The fitted parameters are summarized in Table~\ref{tab:xps_fit_parameters}. The quoted area fractions were calculated relative to the sum of the fitted components within each core-level region and should not be interpreted directly as bulk volume fractions.

For C~$1s$, the lowest-binding-energy component was assigned to
adventitious C--C/C--H species and referenced to 284.8~eV
\cite{Biesinger2022,Skaanvik2025}, corresponding to a shift of
$-1.789$~eV from the as-measured energy scale. The same shift was
applied to the O~$1s$ components. The Mg~$2p$ values are instead
reported on the as-measured scale because the Mg$^{0}$ component
already occurs near the expected metallic binding energy; applying
the C~$1s$-derived shift to the entire Mg~$2p$ spectrum would displace
this metallic component to an unphysical energy. This behavior
suggests differential charging between the conducting Mg layer and
the insulating surface products, consistent with the known
sensitivity of MgO and Mg(OH)$_2$ binding energies to the
charge-referencing procedure \cite{Skaanvik2025}.

\begin{table}[htbp]
  \centering
  \caption{Peak-fitting parameters obtained from the XPS spectra.
  The component assignments were guided by previous XPS analyses
  of magnesium surfaces \cite{Skaanvik2025}. The C~$1s$ and O~$1s$
  energies in the C~$1s$-referenced column were obtained by applying
  a shift of $-1.789$~eV to the as-measured values. The Mg~$2p$
  energies are reported on the as-measured scale; therefore,
  C~$1s$-referenced values are not applicable.}
  \label{tab:xps_fit_parameters}

  \begin{tabular}{llrrrr}
    \hline
    Region & Assignment
      & $E_{\mathrm{B}}^{\text{as-meas.}}$ (eV)
      & $E_{\mathrm{B}}^{\text{C 1s-ref.}}$ (eV)
      & FWHM (eV)
      & Area fraction (\%) \\
    \hline

    Mg~$2p$ & Mg$^{0}$
      & 49.68 & n.a. & 0.72 & 12.9 \\
    Mg~$2p$ & Mg$^{2+}$ (oxide/hydroxide)
      & 51.63 & n.a. & 2.36 & 55.0 \\
    Mg~$2p$ & MgCO$_3$-related
      & 53.63 & n.a. & 3.95 & 32.1 \\
    \hline

    O~$1s$ & MgO lattice oxygen
      & 531.35 & 529.56 & 1.53 & 6.1 \\
    O~$1s$ & Mg(OH)$_2$/hydrated Mg
      & 533.75 & 531.96 & 2.03 & 60.8 \\
    O~$1s$ & MgCO$_3$/hydrated carbonate
      & 535.54 & 533.75 & 2.99 & 33.1 \\
    \hline

    C~$1s$ & C--C/C--H
      & 286.59 & 284.80 & 1.66 & 33.9 \\
    C~$1s$ & C--O/C--OH/C--O--C
      & 288.17 & 286.38 & 1.75 & 15.5 \\
    C~$1s$ & MgCO$_3$/carbonate
      & 291.24 & 289.46 & 3.35 & 50.5 \\
    \hline
  \end{tabular}
\end{table}

The coefficients of determination were $R^2=0.99864$, 0.99940, and 0.99741 for the Mg~$2p$, O~$1s$, and C~$1s$ regions, respectively, with corresponding root-mean-square residuals of 44.4, 223.9, and 102.9 counts. The high-binding-energy Mg~$2p$ component is relatively broad and may include contributions from energy-loss structures and a distribution of local charging shifts in addition to MgCO$_3$. Likewise, the high-binding-energy O~$1s$ component may contain contributions from hydrated carbonate and adsorbed water. The component assignments and area fractions should therefore be regarded as a semiquantitative description of the surface layer rather than as a unique phase analysis.

The O~$1s$ spectrum indicates that the reacted region is
hydroxide-rich rather than pure MgO. Accordingly, a more conservative
summary of the XPS result is an inner MgO/Mg(OH)$_2$-containing layer
of approximately 3--6~nm and an outer carbonate-containing layer of
approximately 1--2~nm. These estimates neglect surface roughness,
lateral inhomogeneity, differential charging, and energy-loss
contributions to the fitted Mg~$2p$ areas and are used only for an
order-of-magnitude comparison with the HHG transmittance.

\subsection{Semiquantitative estimation of the effective reacted-layer thickness}

The effective thickness of the reacted surface region was estimated
using a planar two-region model consisting of a reacted Mg-containing
overlayer and an underlying metallic Mg region. Although the reacted
region contains MgO-, Mg(OH)$_2$-, and carbonate-related species,
their individual depth distributions cannot be determined uniquely
from the present spectra. The Mg~$2p$ components were therefore
grouped into metallic and reacted contributions for the thickness
estimation.

The fitted Mg~$2p$ area fraction assigned to metallic Mg was
$f_{\mathrm{m}}=0.129$, whereas the combined area fraction of the
oxide/hydroxide- and carbonate-related components was
$f_{\mathrm{r}}=0.550+0.321=0.871$. The corresponding
reacted-to-metallic intensity ratio was therefore

\begin{equation}
R_{\mathrm{Mg}}=\frac{I_{\mathrm{r}}}{I_{\mathrm{m}}}=\frac{f_{\mathrm{r}}}{f_{\mathrm{m}}}\simeq 6.75.
\label{eq:mg_intensity_ratio}
\end{equation}

The photoelectron emission angle $\theta$ was defined relative to
the surface normal. For phase $j$, the projected effective
attenuation length was written as

\begin{equation}
L_j=\Lambda_j\cos\theta.
\label{eq:projected_eal}
\end{equation}

Here, $\Lambda_j$ is the practical effective attenuation length
(EAL) of the Mg~$2p$ photoelectrons in phase $j$
\cite{Powell2020EAL,Shard2020}. The measurements
were performed in the normal-emission geometry, such that
$\theta=0^\circ$ and $L_j=\Lambda_j$.

The Mg atomic number density in phase $j$ was calculated from

\begin{equation}
n_{\mathrm{Mg},j}=\frac{\rho_jN_{\mathrm{A}}}{M_j}.
\label{eq:mg_number_density}
\end{equation}

Here, $\rho_j$, $M_j$, and $N_{\mathrm{A}}$ are the mass density,
molar mass, and Avogadro constant, respectively. Each phase
considered here contains one Mg atom per formula unit. The bulk
densities used for Mg, MgO, Mg(OH)$_2$, and MgCO$_3$ were 1.738,
3.58, 2.36, and 2.96~g~cm$^{-3}$, respectively
\cite{Rumble2025CRC}. The MgO value is also consistent with the CXRO
density table \cite{CXRODensity}. The MgCO$_3$ density represents an
anhydrous bulk phase; therefore, these values serve only as nominal
inputs because the actual reacted layer may be hydrated, porous, and
laterally inhomogeneous.

For a reacted overlayer with thickness $d_{\mathrm{r}}$, the
Mg~$2p$ intensity originating from the reacted layer was expressed
as

\begin{equation}
I_{\mathrm{r}}=CS_{\mathrm{Mg}\,2p}n_{\mathrm{r}}L_{\mathrm{r}}\left[1-\exp\left(-\frac{d_{\mathrm{r}}}{L_{\mathrm{r}}}\right)\right].
\label{eq:reacted_intensity}
\end{equation}

The intensity from the underlying metallic Mg region was written as

\begin{equation}
I_{\mathrm{m}}=CS_{\mathrm{Mg}\,2p}n_{\mathrm{m}}L_{\mathrm{m}}\exp\left(-\frac{d_{\mathrm{r}}}{L_{\mathrm{r}}}\right).
\label{eq:metallic_intensity}
\end{equation}

Here, $C$ contains the X-ray flux and analyzer transmission, and
$S_{\mathrm{Mg}\,2p}$ is the Mg~$2p$ sensitivity factor.
Combining Eqs.~(\ref{eq:reacted_intensity}) and
(\ref{eq:metallic_intensity}) gives

\begin{equation}
R_{\mathrm{Mg}}=\frac{n_{\mathrm{r}}L_{\mathrm{r}}}{n_{\mathrm{m}}L_{\mathrm{m}}}\left[\exp\left(\frac{d_{\mathrm{r}}}{L_{\mathrm{r}}}\right)-1\right].
\label{eq:overlayer_ratio}
\end{equation}

The effective reacted-layer thickness was therefore obtained from

\begin{equation}
d_{\mathrm{r}}=L_{\mathrm{r}}\ln\left[1+R_{\mathrm{Mg}}\frac{n_{\mathrm{m}}L_{\mathrm{m}}}{n_{\mathrm{r}}L_{\mathrm{r}}}\right].
\label{eq:reacted_thickness}
\end{equation}

Because both contributions were obtained from the same Mg~$2p$
core level and their kinetic energies differ by only a few
electronvolts, the Mg~$2p$ photoionization cross section, analyzer
transmission, and relative sensitivity factor cancel to a good
approximation in the intensity ratio.

For Al~K$\alpha$ excitation, the kinetic energy of the Mg~$2p$
photoelectrons was approximately 1433--1437~eV. The IMFPs calculated
using the TPP-2M predictive equation at this energy are of the order
of 3--4~nm \cite{Tanuma1994}. For the thickness
estimate, practical EALs of 2--3~nm were adopted to account
approximately for elastic-scattering effects and uncertainty in
the composition of the reacted region \cite{Powell2020EAL}. The Mg
atomic number
density of metallic Mg was approximately
$n_{\mathrm{m}}=4.3\times10^{22}$~cm$^{-3}$.

The O~$1s$ component areas assigned to MgO, Mg(OH)$_2$, and
carbonate-related oxygen were used to construct a nominal
composition for the optical calculations. The areas were corrected
for the number of oxygen atoms per Mg atom and for the molar volume
of each phase. The nominal volume fraction $\phi_i$ of phase $i$
was calculated from

\begin{equation}
\phi_i=\frac{(A_i/\nu_i)(M_i/\rho_i)}{\sum_k(A_k/\nu_k)(M_k/\rho_k)}.
\label{eq:nominal_volume_fraction}
\end{equation}

Here, $A_i$ is the O~$1s$ component area and $\nu_i$ is the number
of oxygen atoms per Mg atom. The resulting nominal volume fractions
and the component thicknesses used in the optical calculations are
summarized in Table~\ref{tab:reacted_layer_model}.

\begin{table}[t]
\centering
\caption{Nominal composition of the reacted surface region and
component thicknesses used in the optical calculations.}
\label{tab:reacted_layer_model}
\begin{tabular}{lccccc}
\hline
Phase & $A_i$ & $\nu_i$ & $\phi_i$ &
$d_i$ (4 nm) & $d_i$ (8 nm) \\
\hline
MgO
& 0.061 & 1 & 0.060 & 0.24 nm & 0.48 nm \\
Mg(OH)$_2$
& 0.608 & 2 & 0.663 & 2.65 nm & 5.30 nm \\
MgCO$_3$
& 0.331 & 3 & 0.277 & 1.11 nm & 2.22 nm \\
\hline
Total
& 1.000 & -- & 1.000 & 4.00 nm & 8.00 nm \\
\hline
\end{tabular}
\end{table}

The average Mg atomic number density of the nominal reacted region
was calculated as
$n_{\mathrm{r}}=\sum_i\phi_i\rho_iN_{\mathrm{A}}/M_i$ and was
approximately $2.5\times10^{22}$~cm$^{-3}$. Using this nominal
composition in Eq.~(\ref{eq:reacted_thickness}) gives a
reacted-layer thickness of approximately 5--8~nm for EALs between
2 and 3~nm.

To account for the additional uncertainty in composition, limiting
number densities corresponding to MgO-rich and
hydroxide/carbonate-rich reacted regions were also considered.
Combining these composition limits with the adopted EAL range gives
an overall effective reacted-layer thickness of approximately
4--8~nm.

The lower and upper bounds of 4 and 8~nm were subsequently used in
the optical sensitivity calculations. For simplicity, the total
geometrical thickness of the Mg-derived region was fixed at
200~nm. The remaining metallic Mg thicknesses were therefore
196 and 192~nm for the 4- and 8-nm models, respectively.

These estimates are model-dependent and should not be interpreted
as statistical confidence intervals or as a unique depth profile.
In particular, attenuation of the inner components, overlap among
the Mg$^{2+}$ Mg~$2p$ signals, differential charging, and possible
contributions from energy-loss structures introduce additional
uncertainty. The 4--8~nm range was therefore used only to evaluate
the sensitivity of the calculated XUV transmittance to the reacted
surface region.

\section{Layered optical model of the Mg film}
\label{sec:optical_model}

The Mg-film transmittance was evaluated using an ideal-film model
and two XPS-informed layered models. The total geometrical thickness
of the Mg-derived layer was fixed at 200~nm. The reacted region was
modeled using nominal volume fractions of 0.0600 MgO, 0.6625
Mg(OH)$_2$, and 0.2775 MgCO$_3$.

For a reacted-layer thickness $d_{\mathrm{r}}$, the metallic-Mg
thickness was set to $200~\mathrm{nm}-d_{\mathrm{r}}$, and the
thickness of each reacted component was calculated as
$\phi_j d_{\mathrm{r}}$, where $\phi_j$ is its nominal volume
fraction. Thus, the 4-nm model consisted of 196~nm of metallic Mg,
0.24~nm of MgO, 2.65~nm of Mg(OH)$_2$, and 1.11~nm of MgCO$_3$.
The corresponding thicknesses for the 8-nm model were 192, 0.48,
5.30, and 2.22~nm, respectively. A 100-nm-thick Parylene-N
supporting membrane was included in all calculations.
\begin{figure*}[t]
    \centering    \includegraphics[width=0.9\linewidth]{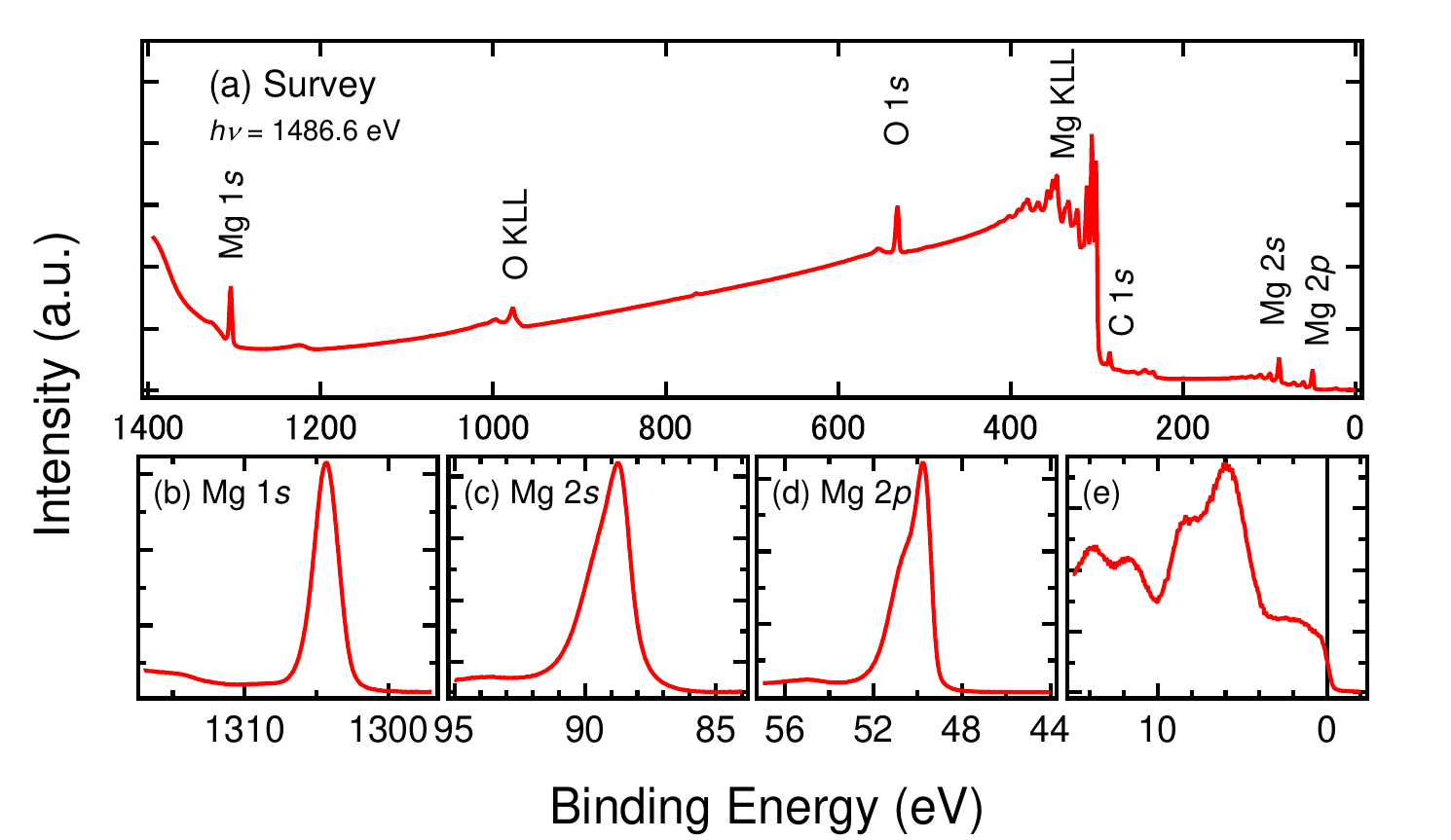}
    \caption{\justifying
        X-ray photoelectron spectroscopy (XPS) spectra of the sample.
        (a) Survey spectrum and high-resolution spectra of
        (b) Mg~1$s$, (c) Mg~2$s$, (d) Mg~2$p$, and
        (e) the valence-band region near the Fermi level. The measurements were performed using Al~K$\alpha$ radiation
        with a photon energy of $h\nu = 1486.6$~eV.}
    \label{fig:XPS_Mg}
\end{figure*}
The total transmittance was calculated as
\begin{equation}
\begin{split}
T_{\mathrm{total}}(E;d_{\mathrm{r}})
={}&
T_{\mathrm{Parylene}}(E;100~\mathrm{nm})\\
&\times
T_{\mathrm{Mg}}
  \left(E;200~\mathrm{nm}-d_{\mathrm{r}}\right)
\prod_j T_j(E;\phi_jd_{\mathrm{r}}),
\end{split}
\label{eq:optical_model}
\end{equation}
where $d_{\mathrm{r}}=0$, 4, and 8~nm correspond to the ideal,
4-nm reacted-layer, and 8-nm reacted-layer models, respectively,
and $j$ denotes MgO, Mg(OH)$_2$, and MgCO$_3$. The transmittance
of each constituent was calculated using the optical constants
tabulated in the CXRO database \cite{Henke1993}. No parameters were
adjusted to reproduce the measured HHG transmittance.

The calculated models should therefore be regarded as a sensitivity
analysis of the influence of the reacted surface region rather than
as a quantitative reproduction of the measured transmittance.


\bibliography{references}